# The growth of 2D crystalline g-$C_3N_4$ films and the control of optoelectronic properties

Ying Du[1], Meng Wu[1*], Hui-Qiong Wang[1*,2], Junyong Kang[1]

1. Engineering Research Center of Micro-nano Optoelectronic Materials and Devices, Ministry of Education; Fujian Key Laboratory of Semiconductor Materials and Applications, CI Center for OSED, and Department of Physics, Xiamen University, Xiamen 361005, P.R. China
2. Xiamen University Malaysia, Sepang 43900, Selangor, Malaysia

*Corresponding authors. Email:

memg.wu@xmu.edu.cn; hqwang@xmu.edu.cn

## ABSTRACT

g-C3N4 is a novel semiconductor photocatalyst material; however, the low specific surface area and rapid carrier compliance hinder its photocatalytic performance. On the other hand, the synthesis of 2D g-C3N4 with high crystallinity remains challenging. Here, we report the growth of 2D crystalline g-C3N4 films with thicknesses up to 100 nm on the indium tin oxide substrates by chemical vapor deposition. The films show high quality, as shown by scanning electron microscopy and X-ray diffraction, and exhibit intense fluorescence at room temperature. The optimal growth conditions, such as temperature and carrier gas flow rate, were achieved by analyzing their effects on the electronic structure through X-ray absorption spectra and X-ray photoelectron spectroscopy. By adding thiourea to the melamine precursors, we introduced N vacancies to achieve band gap modulation and promote carrier separation. This work provides guidelines for the further improvement of g-C3N4 performance and for extending its application in the field of photocatalytic devices.

## 1. INTRODUCTION

The introduction of polymeric semiconductor materials has broken through the bottleneck in the development of traditional semiconductor materials and laid the foundation for the synthesis of low-cost, application-flexible devices, becoming the cornerstone of a new generation of semiconductor devices.[1] With its excellent physicochemical stability and unique electronic properties, g-$C_3N_4$ stands out among many polymeric semiconductors and has become a candidate for functional materials for optoelectronic applications in recent years.[2] Thin film deposition remains as one of the core techniques in optoelectronic devices, thus the growth of g-$C_3N_4$ thin films is a necessary path for its development towards its promising applications for post-silicon electronics.[3]

Researchers have tried various methods to synthesize g-$C_3N_4$ films since 2015.[4,5] They can be classified as the top-down methods similar to spin-coating, in which, the g-$C_3N_4$ powder is dissolved and applied on the substrate,[6] and the bottom-up methods such as electrophoretic deposition and electrostatic spinning techniques, in which, it is polymerized directly on the substrate.[7–9] The g-$C_3N_4$ synthesized by the top-down methods are mostly incomplete and coarse since defects arising from the dissolved powders and volatilization of solvents are inevitable during film formation. In contrast, g-$C_3N_4$ films yielded by the bottom-up methods exhibit a relatively uniform and controllable thickness mostly in the micron range. Therefore, obtaining two-dimensional films with high crystallinity remains a challenge, which is important both for improving its specific surface area and carrier transport efficiency and facilitating the fundamental physical properties as well as its application in optoelectronic devices. Among the many growth methods, chemical vapor deposition (CVD) is the most promising method for growing large-area high-quality g-$C_3N_4$ films.[10,11] Since the successful growth of graphene crystals by CVD, the method has been extended for the synthesis of other 2D materials, such as various 2D transition metal dichalcogenide materials, and achieved the growth of large single crystals.[12] Here, we have grown high-quality g-$C_3N_4$ thin films by CVD using melamine as a precursor. Prior to this, several teams reported the synthesis of g-$C_3N_4$ films by CVD, and this breakthrough is of great significance for the development of g-$C_3N_4$. In 2020, Giusto et al. for the first time grew g-$C_3N_4$ films with high crystallinity by CVD, illustrating the feasibility of this method.[2] Chen et al. deposited ultrathin films with poor crystallinity by CVD[13], and Chubenko et al. optimized the method to improve the crystallinity of the films, but the films were thicker and less uneven. [14] However, the influence of growth conditions on the crystal structure and electronic structure of g-$C_3N_4$ thin films during growth has been lacking a strong explanation, which is crucial to clarify the growth process of crystalline g-$C_3N_4$ thin films, and it is a strong guide for the growth of the material, which is important for the optimization of its subsequent properties and the application on the re-device. Meanwhile, defect engineering is an effective measure to improve optoelectronic performance, but most of the existing research is based on complex experimental processes.[15–17] We simplified the defect engineering and established the basis for the research of its electronic structure and the further improvement of its photocatalytic performance.

## 2. EXPERIMENTAL SECTION

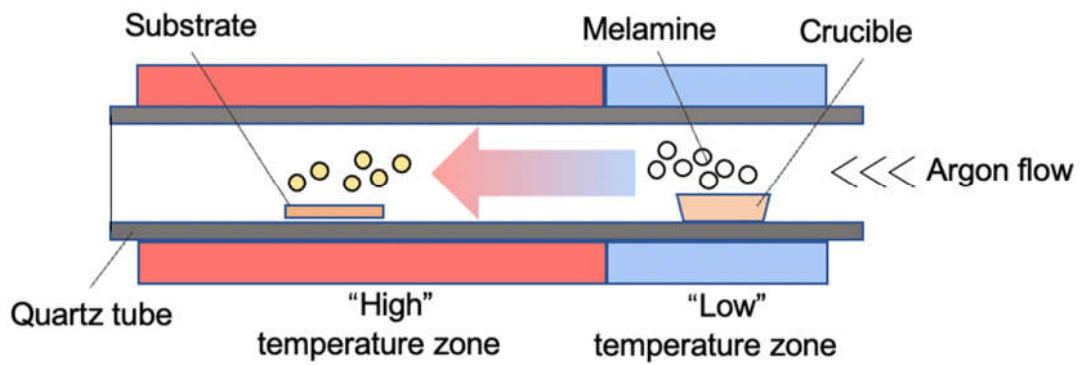

**Figure 1.** Schematic view of the tube furnace for chemical vapor deposition of g-$C_3N_4$ thin films.

## 2.1 Preparation of g-C₃N₄ crystalline films

A 10mm*10mm double-sided polished indium tin oxide (ITO) glass was chosen as the substrate, which was cleaned in an ultrasonic bath with acetone, anhydrous ethanol, and deionized water, separately for five minutes, and transferred into the high-temperature zone of the quartz tube (Fig.1). The 2g of melamine or the mixture precursors of melamine and thiourea (2 g total) in the ratios of 1:1 and 1:3, which were mechanically ground, was taken into the open quartz boat and then put into the low-temperature zone of the tube furnace. Before starting the experiment, tube furnace was pumped down to $1*10^{-1}$ pa, then the tube furnace was filled with argon. The process was repeated twice, and the argon flow was set in the range of 100SCCM-200SCCM. The "high-temperature zone" where the substrate was located was first heated up to 550°C, and then the "low-temperature zone" where the precursor was located started to heat up to 350 °C. After 100 mins, the heating was stopped and the tube furnace was cooled down to room temperature under argon protection.

## 2.2 Structural characterization methods

Scanning Electron Microscopy (SEM, ZEISS SIGMA-HD) was used to measure the microscopic morphology of the sample. X-ray diffraction (XRD, Rigaku Ultima IV) was conducted to characterize the crystal structure of thin films. X-ray absorption spectroscopy (XAS) measurements were performed at the 4B9B beamline of the Beijing Synchrotron Radiation Facility (BSRF) to characterize the atomic structure of the sample. The UV−Vis diffuse reflectance spectra were obtained by using a Scan UV−vis spectrophotometer (LAMBDA 850). Photoluminescence (PL, Edinburgh LFS1000) spectra were measured with a fluorescence spectrophotometer. X-ray photoelectron spectroscopy (XPS, ESCALAB Xi+) was performed to analyze the content of elements. Valence Band (VB)-XPS was performed using ESCALAB Xi+ electronic energy analyzer

## 3. RESULTS AND DISCUSSION

### 3.1 Synthesis of g-C$_3$N$_4$ thin films

There have been numerous researches on the effect of substrate temperature on g-C$_3$N$_4$ thin film deposition using CVD method, but the choice of carrier gas and its flow rate, as well as the distance between substrate and precursor, are also important parameters.

### Growth temperature

We have determined the optimal growth temperature by SEM observation of the sample pattern at different temperatures in the 500-600°C temperature range of the substrate. (Fig.2(a-d)) We found that samples grown at 550℃ had the highest deposition rate and the flattest surface. In this section we will respectively discuss the specific effects of the carrier gas flow rate and the distance of the precursor from the substrate on its structure during the growth of g-C$_3$N$_4$ films.

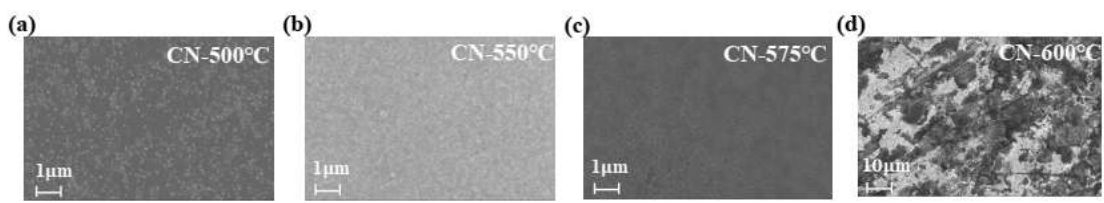

**Figure 2.** SEM diagram of sample surface when the substrate temperature is 500°C (a), 550°C (b), 575°C (c) and 600°C (d).

### Carrier gas flow rate

Argon, a chemically inert gas, has been chosen as the carrier gas, which can not only bring the precursor to the substrate accompanying with the gas flow, but also prevent the material from being oxidized by the trace oxygen in the gas mixture. Typically, the transport rate of the precursor to the substrate can be controlled by the flow rate of the carrier gas, where different flow rates of 100SCCM, 150SCCM and 200SCCM are considered, respectively (Fig.3(a)).

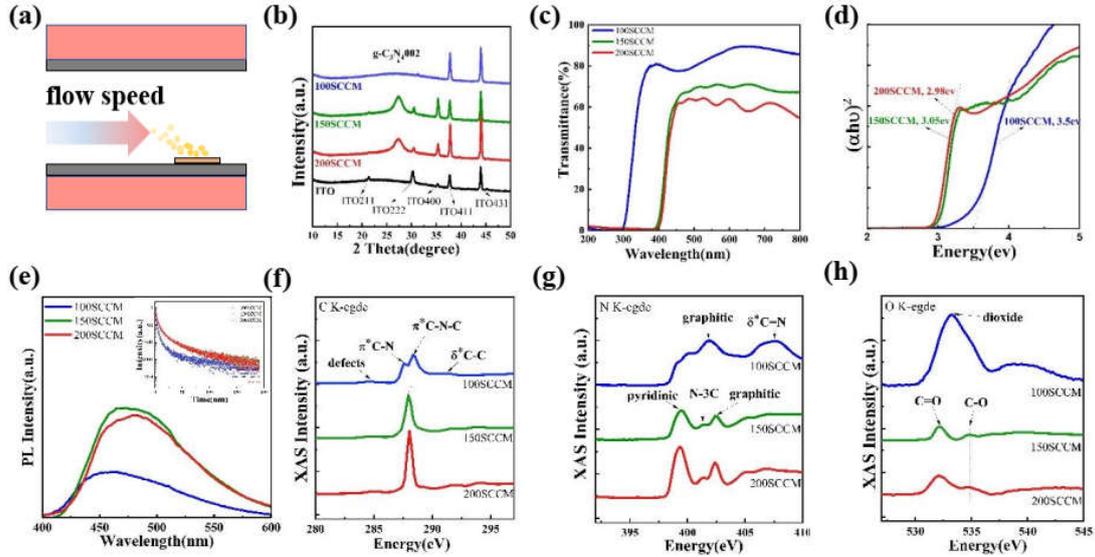

**Figure 3.** (a) Schematic illustration of the flow rate of carrier gas during CVD deposition. (b) XRD patterns of g-$C_3N_4$ films at different carrier gas flow rates. (c) The transmittance spectra, the band gap of g-$C_3N_4$ obtained from UV-vis absorption spectra (d) of g-$C_3N_4$ films. PL patterns of g-$C_3N_4$ films (e) and inset shows the fluorescence lifetime. (f-h) XAS spectra at C K-edge, N K-edge and O K-edge of the reacting g-$C_3N_4$ films.

Figure 3(b) shows the XRD scans of g-$C_3N_4$ films on ITO substrates where good crystallinities are obtained at flow speed of 150 SCCM and 200 SCCM, with distinct (002) diffraction peak. At the flow speed of 100 SCCM, no g-$C_3N_4$ characteristic peaks were observed. Based on the transmission spectra (Fig.3(c)), it was found that the transmittance of the films grown at 100SCCM was higher than 80%, thus it was speculated that the lower flow rate of 100SCCM led to lower air pressure inside the tube, which affected the deposition mechanism of the films and led to a decrease in the densities of the films with more defects. By comparing different flow speed transmission spectra (Fig.3(c)) we found that the transmission of the films decreases with increasing flow rates, which is attributed to the thickening of the films. In addition, it can be observed that there are interference fringes in the transmission spectra of 150SCCM and 200SCCM, which indicates that the optical constants remain constant when the films are irradiated with photons due to the uniformity of the films[18], which also indicates the high quality of the films. The corresponding change of bandgap as a function of flow rate is illustrated in Fig. 3(d).

In the PL spectra (Fig.3(e)), a redshift appears as the wavelength gets longer, which corresponds to its UV spectrum. The lower fluorescence intensity of 100SCCM is due to its more defective states, which promotes non-radiative charge carrier relaxation. The enhanced carrier gas flow rate results in a significant enhancement of the fluorescence intensity attributed to the further coalescence of its Π-conjugated system at high air pressure,[19] which leads to a further increase in crystallinity. As shown in the fluorescence lifetime, the 150SCCM flow rate exhibits the strongest fluorescence lifetime of 12.428 ns. However, the current fluorescence lifetime of most g-$C_3N_4$ is below 10 ns,[20][21][22] thus we exceeded most previous studies, further demonstrating the weak non-radiative jump and energy loss under this condition.

Figure 3(f-h) exhibits the experimentally measured XAS spectra at the C K-edge, the N K-edge and the O K-edge. As shown in Fig. 3(f), the standard g-$C_3N_4$ specific C-N-C Π* resonance at 288.0 eV is shown at both flow rates of 200 SCCM and 150 SCCM. While at a flow rate of 100 SCCM, a weak shoulder at 287.55eV is assigned to the Π* resonance of C-N species.[23] Meanwhile, a relatively obvious δ* peak appears at 291.2ev. It can be explained that the C-N-C structure is not fully formed at the flow rate of 100 SCCM. As shown in Fig. 3(g) N K-edge, the flow rate of 150SCCM and 200SCCM shows the typical characteristic peak of N in g-$C_3N_4$. At 100 SCCM, the pyridinic N peak is very significantly weakened and a broadened δ* peak appears at 407.4 ev. It can be assumed that the C=N specific gravity of the films obtained at a flow rate of 100 SCCM increases and does not form a complete heptazine unit. [24]

By analyzing the K-edge of C, N, and O elements according to XAS, the results consistently show that 150SCCM and 200 SCCM flow rates are suitable for g-$C_3N_4$ film growth, while 100 SCCM flow rate is not sufficient to obtain g-$C_3N_4$ films with complete structures.

### Source–substrate distance

The distance between the precursor and the substrate (Fig. 4(a)) is of great importance for the film growth. Here we choose three points with the same temperature in the high-temperature zone of the tube furnace to place the ITO substrate with the same treatment. The samples were grown at a temperature of 550 ℃ under a flow rate of 150 SCCM argon gas.

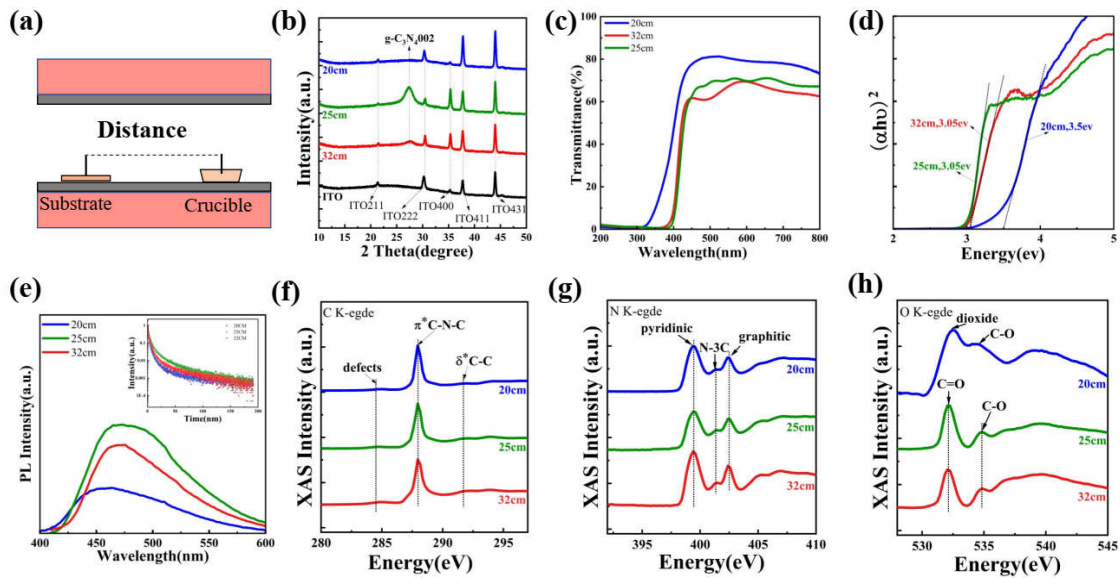

**Figure 4.** (a) Schematic illustration of the experiments of source–substrate distance. (b) XRD pattern of g-$C_3N_4$ films at different source–substrate distance. (c) Schematic of the transmittance of g-$C_3N_4$ films. (d) Schematic of the band gap of g-$C_3N_4$ obtained from UV-vis absorption spectra. (e) PL patterns of g-$C_3N_4$ films and the fluorescence lifetime. (f-h) Schematic of XAS of the atomic structure of the reacting g-$C_3N_4$ films.

The XRD spectra (Fig.4(b)) shows the sharpest g-$C_3N_4$ peak at 25 cm, and this peak is obviously influenced by the precursor–substrate distance. As the distance of the substrate from the precursor becomes farther, this peak shows a trend of becoming stronger and then weaker. As observed by SEM, the surface of the sample was found to be incomplete with more cracks at a distance of 20 cm (Fig.5(c)), while the surface is smooth and uniform at a distance of 25 cm (Fig.5(b)) and a grainy surface at 32 cm (Fig.5(a)). Thus, it can be speculated that the film deposition becomes more uniform and the crystallinity is improved as the distance gets farther, while the film integrity is affected due to the low concentration of precursors as the distance continues to get farther. Based on UV-vis spectra (Fig.4(c)), it is found that the sample grown at 20 cm from the substrate exhibits ultra-high transmittance and a wide band gap attributable to the presence of defects on its surface.

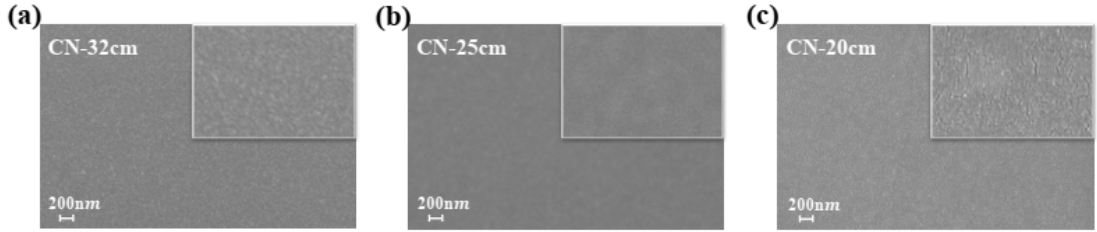

**Figure 5.** SEM schematic of g-C$_3$N$_4$ film surface at the source–substrate distance of 32 cm (a), 25cm (b) and 20cm (c).

In order to investigate the effect of distance on the optical properties of the samples, we analyzed the PL spectra (Fig.4(e)) and found that the 25 cm film had the strongest fluorescence and the TRPL showed the longest fluorescence lifetime, reaffirming that 25 cm is the best location for sample growth.

To further investigate the change in the atomic structure behind the different optical properties of the material, an analysis of the atomic structure of g-C$_3$N$_4$ at different positions was performed based on XAS spectra (Fig.4(f-h)). At a distance of 20 cm, the K-edge of oxygen shows a clear change. The elemental ratio of XAS to XPS clearly shows an increase in the content of oxygen and a decrease in the content of nitrogen. Thus, it is considered that the trace amount of oxygen in the environment replaces the position of nitrogen to form covalent bonds with C atoms, thus leading to the incomplete structure of g-C$_3$N$_4$.[25]

## Intrinsic properties of g-C$_3$N$_4$ with optimized deposition conditions

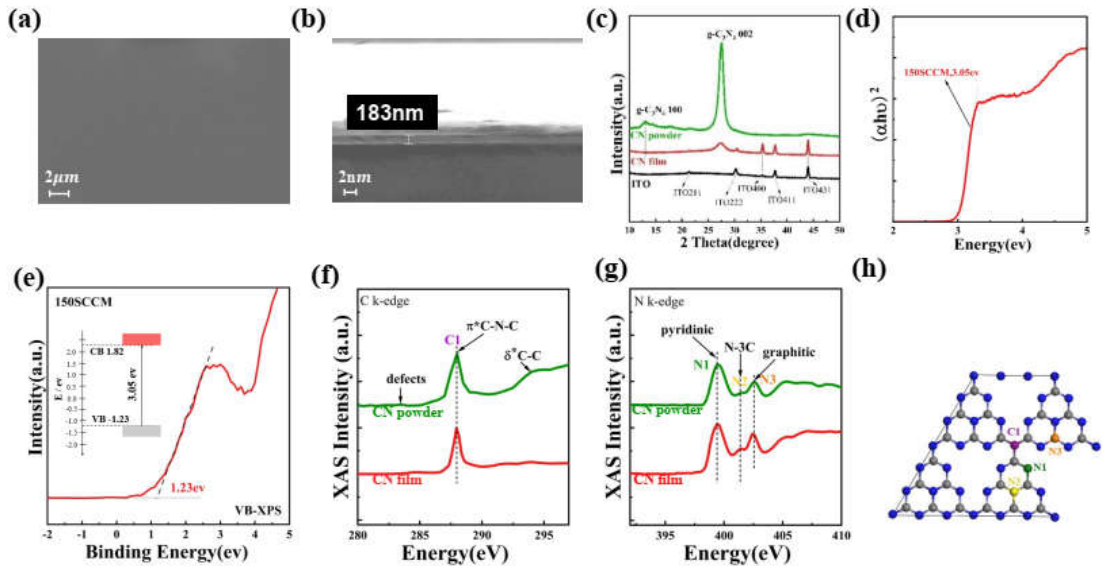

**Figure 6.** (a) SEM schematic of the surface samples of g-$C_3N_4$ films. Panel(b) shows the sample thickness schematic obtained from the SEM image of the sample section. (c) XRD patterns of g-$C_3N_4$ film and powder. (d) Schematic diagram of the energy band gap of g-$C_3N_4$ film. (e) shows the valence band diagram of the g-$C_3N_4$ film, and the inset shows its band gap structure. (f-g) XAS spectra at C K-edge, N K-edge of the g-$C_3N_4$ film and powder. (h) molecular structure

Following the above discussion, we determined the optimal conditions for the growth of the films at a substrate to precursor distance of 25 cm, the substrate at 550 °C, and the argon atmosphere with 150 SCCM flow rate. The intrinsic properties of g-$C_3N_4$ thin films obtained with the optimized deposition conditions will be discussed in the section. Figure 6(a-b) shows the cross section SEM image of g-$C_3N_4$ films deposited on ITO substrate, which exhibits a continuous and highly homogeneous structural morphology with a thickness of ~180 nm. Figure 6(c) compares the XRD spectra of the g-$C_3N_4$ film and powder; the peak at 27.53° corresponds to the (002) plane of g-$C_3N_4$,[4] which indicates the structure of g-$C_3N_4$ as interlayer stacking with an interlayer distance of about 0.32 nm.[22] A peak at 12.98° in the powder sample corresponds to a typical reflection in the 100 plane, indicating a structural inter-pore distance of 0.675 nm in the g-$C_3N_4$ layers. The peak is absent in g-$C_3N_4$ thin film, which is due to the lower intensity in g-$C_3N_4$ with nm thickness.[14] Figure 6 (d) shows the band gap of the g-$C_3N_4$ film, which is estimated as 3.05ev from the UV spectra. The valence band maximum is located at -1.23eV relative to the Fermi level, according to the VB-XPS measurement (Fig.6(e)), and the conduction band minimum is +1.77eV as determined by the equation $E_{CB} = E_{VB} – E_g$.[26]

We have further confirmed the electronic structure by XAS analyses. As shown in Fig. 6 (f-g), the XAS spectrum at the C K-edge has a clear Π* resonance at 288.0 eV as a characteristic resonance of C in g-$C_3N_4$, attributed to C-N-C.[27] According to previous research, the C defect peak at 284.6 eV may be caused by the combination of C with a heteroatom such as O.[28] As shown in the XAS spectra of the N K-edge, the two distinct Π* peaks at 399.4 eV and 402.4 eV correspond to the pyridinic and graphitic N species.[24] The weaker Π* peak in the middle corresponds to the N-3C in the middle of the heptazine ring. The above XAS analysis of g-$C_3N_4$ films and comparison with g-$C_3N_4$ powder reveals that the structure of g-$C_3N_4$ films is more complete and there are almost no defect peaks and other bonding peaks, which further verifies the high purity and structural integrity of our g-$C_3N_4$ films..[29]

## 3.2 Synthesis of g-$C_3N_4$ films with N vacancies

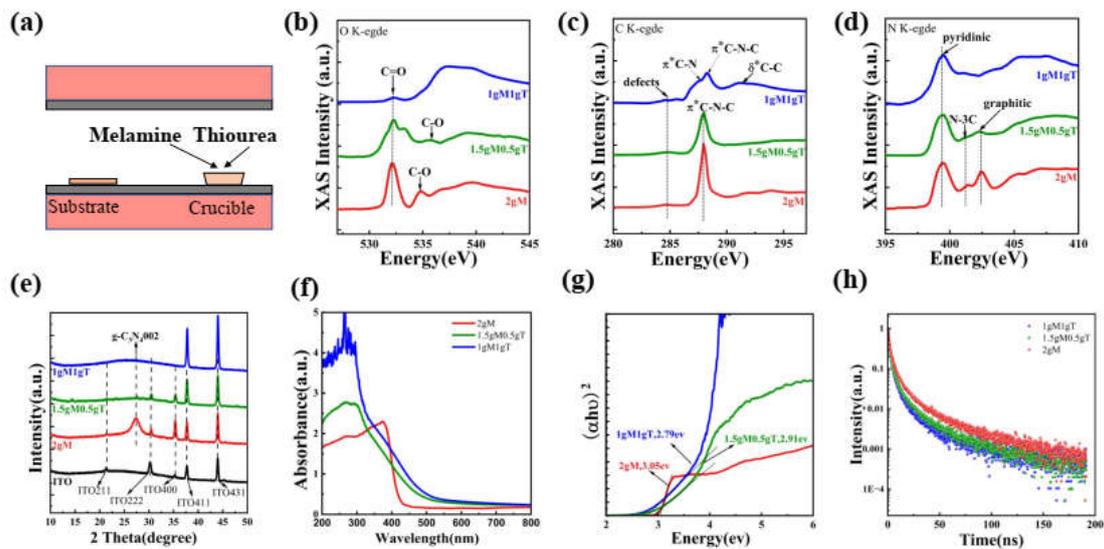

**Figure 7.** (a) Schematic illustration of the experiments of source–substrate distance. (b-d) XAS patterns of g-$C_3N_4$ films of 2gM, 1.5gM0.5gT and 1gM1gT precursors. (e)

XRD patterns of precursors of 2gM, 1.5gM0.5gT, and 1gM1gT. (f) Schematic illustration of UV-Vis absorption spectra of g-$C_3N_4$ films. (f)Schematic illustration of the band gap of g-$C_3N_4$ obtained from UV-vis absorption spectra. (g)Schematic illustration of band gap structure of g-$C_3N_4$ films. (h) Diagram of fluorescence lifetime of g-$C_3N_4$ films.

Under optimal growth conditions, we achieved a controlled introduction of nitrogen vacancies in g-$C_3N_4$ films by doping thiourea in melamine precursors. XRD spectra (Fig.7(e)) were analyzed for the crystal structure of g-$C_3N_4$ obtained from different precursors. With the increase of the proportion of thiourea added to the melamine precursors, the strong peak at 27.53° corresponds to the g-$C_3N_4$ interlayer stacking weakened to disappear. On the one hand, it is considered that the addition of thiourea thins the layer thickness of g-$C_3N_4$,[30] which is consistent with our previous research on g-$C_3N_4$ powder. On the other hand, it is considered that the addition of thiourea makes the complete heptazine structure defective.

To further verify this speculation, we performed an elemental analysis based on XPS. Table 1 shows that the content ratio of C/N (wt%) was increased from 0.85 (the 2g melamine) to 1.27 (the 1.5g melamine and 0.5g thiourea), and the atomic content of oxygen was significantly increased. This is explained by the increased loss of nitrogen after the incorporation of thiourea in the melamine precursors and the doping of surface oxygen elements, which makes the samples have more defects.[31] This conclusion was also verified in the XAS spectra(Fig.7(b-d)). The spectrum has no peak assigned to the S species, implying that the S in thiourea is released as a gas during the counter reaction.[22] While when 1 g of thiourea was incorporated in the precursor, the N content of the obtained sample was only 6.32%, S content of 16.84%. The analysis combined with the XAS characterization leads to only a small amount of pyridinic N remains in the film, which is considered as the excess thiourea in the precursor inhibits the reaction, making the triazine structure with more N vacancies, while the precursor reaction is incomplete.

| Sample | C(%) | N(%) | S(%) | O(%) |
|---|---|---|---|---|
| 2gM | 45.38 | 53.50 | / | 1.12 |
| 1.5gM0.5gT | 46.68 | 36.76 | 3.62 | 12.93 |
| 1gM1gT | 53.04 | 6.84 | 20.43 | 19.69 |

**Table 1**. Elemental analysis of CN-2gM, CN-1.5gM0.5gT and CN-1gM1gT samples.

By analyzing the atomic structure, we demonstrate that the addition of thiourea to the precursor leads to the generation of N vacancies in the obtained films, which helps to suppress the carrier compliance and improve the utilization of visible light.[30] To further investigate the effect of N vacancies on their properties, we performed a series of characterizations of the optoelectronic properties.

We obtained its band gap based on UV-vis absorption spectra (Fig.7(f-g)) indicating that the band gap narrows with increasing proportion of thiourea in the precursor. When 0.5g of thiourea was added to the precursor, the band gap was reduced from 3.05ev to 2.91ev, and the band gap continued to be reduced to 2.79ev when 1g of thiourea was added to the precursor. This is attributed to N vacancies inducing electron leap to additional energy levels causing a reduction in the band gap, thus allowing a fuller utilization of visible light in the field of photocatalysis.[32] Meanwhile, the addition of thiourea accelerates the decay of the fluorescence lifetime of the film (Fig.7(h)), which also indicates that the N-defective sites effectively trap electrons and thus inhibit the carrier complex. [33]

## CONCLUSION

In summary, we obtained high-quality g-$C_3N_4$ films by CVD using melamine as the precursor, which exhibited excellent optoelectronic properties. And we used XAS and other characterizations to analyze its atomic structure and photovoltaic properties, detailing the significant effects of carrier gas flow rate and source–substrate distance on its crystallization during the growth process. In addition, we have realized the controlled introduction of N vacancies and further modulation of the atomic and electronic structures by a simple operation of incorporating different contents of thiourea into the melamine precursors. These research results are of great significance for the growth of highly crystalline g-$C_3N_4$ thin films and provide a new idea for defect engineering in g-$C_3N_4$ thin films, laying the foundation for future applications in photocatalysis and other fields.